\documentclass[twocolumn,secnumarabic,amssymb,nobibnotes,aps,prl]{revtex4-1}
\usepackage{graphicx}
\usepackage{textcomp}
\usepackage{amsmath}
\usepackage{commath}
\usepackage{color}
\newcommand{\sg}[1]{}
\renewcommand{\sg}[1]{{\color{blue}{#1}}} 

\begin{document}
	\title{Soft topological lattice wheels}
	\author{William Zunker}
	\author{Stefano Gonella}
	\email{sgonella@umn.edu}
	\affiliation{{Department of Civil, Environmental, and Geo- Engineering}\\ University of Minnesota, Minneapolis, MN 55455, USA\\}
	
	\begin{abstract}
		 We investigate the large deformation and extreme load-management capabilities of a soft topologically polarized kagome lattice mapped to a cylindrical domain through the problem of a lattice wheel rolling on an irregular surface. We test the surface-lattice interaction experimentally by subjecting a 3D-printed topological lattice wheel prototype to localized and distributed boundary loads. This investigation reveals a dichotomy in the force transfer between the two loading scenarios, whereby localized loads are absorbed with limited stress penetration into the bulk and small force transfer to the wheel axle, compared to distributed loads. Through numerical simulations, we compare the lattice wheel against a baseline solid wheel to highlight the unique stress management opportunities offered by the lattice configuration. These findings promote the design of rolling objects enabled by topological mechanics, in which a surplus of softness, activated by local asperities, can coexist with a globally stiff response to distributed loads that ensures satisfactory load-bearing capabilities. 
		 \bigskip
		 
		\vspace{0.4cm}
	\end{abstract}
		 
	\maketitle
	
	\section{Introduction}
	
	Mechanical metamaterials~\cite{bertoldi2017flexible} are architected materials, typically involving a periodic tessellation of a unit cell, designed to exhibit unconventional response to a variety of external loads. The metamaterial design philosophy, in which the geometry, rather than the material properties, dictates the mechanical response, has resulted in numerous engineering innovations, including structural materials with zero or negative effective mechanical properties~\cite{nicolaou2012mechanical, babaee20133d, buckmann2014elasto, yasuda2015reentrant}, shape morphing and programmable capabilities~\cite{coulais2016combinatorial, haghpanah2016multistable, celli2018shape, jin2020kirigami}, nonreciprocal behavior~\cite{coulais2017static}, asymmetric edge behavior~\cite{kane2014topological, chen2014nonlinear, rocklin2017transformable}, and robustness against defects and disorder~\cite{khanikaev2015topologically, mousavi2015topologically, wang2015topological, pal2016helical}.
	
	A number of recent advances in mechanical metamaterials has spurred from the field of topological mechanics~\cite{kane2014topological, chen2014nonlinear, prodan2009topological, paulose2015topological, paulose2015selective, nash2015topological, khanikaev2015topologically, susstrunk2015observation, mousavi2015topologically, wang2015topological, meeussen2016geared, rocklin2016mechanical, chen2016topological, pal2016helical, rocklin2017transformable,  zhang2018fracturing, zhou2020switchable, sun2020continuum}, which has arisen from the contamination of classical mechanics problems with concepts of topological physics~\cite{haldane1988model, kane2005quantum, hasan2010colloquium, qi2011topological}. This crossover of concepts has led to mechanical systems with properties dictated by their topology that are intrinsic to the bulk but manifest at the edges and or at internal interfaces, according to the so-called bulk-edge correspondence. Furthermore, these properties are preserved as long as the topological phase of a system is preserved, leading to robustness against perturbations in the geometry, a notion referred to as topological protection. Pertinent to the work presented in this Letter is a subset of topological phenomena arising in Maxwell lattices. In the context of ideal systems made of struts connected by ideal hinges~\cite{maxwell1864calculation}, two-dimensional Maxwell lattices are characterized by a coordination number equal to 4. This places them between under-coordinated lattices, capable of supporting a large number of floppy modes (i.e. zero-frequency mechanisms~\cite{mao2018maxwell}) in the bulk, and over-coordinated lattices, which are stiff. In finite Maxwell lattices under open boundary conditions, a problem of great interest is the existence of floppy modes localized at the edges~\cite{sun2012surface}. Kane and Lubensky showed that certain lattices exhibit topological polarization, whereby the zero modes are focused on a specific edge, deemed the floppy edge, leaving the other edge(s) stiff~\cite{kane2014topological}. Similar topological attributes have been shown and exploited for structural lattices with non ideal hinges to harness asymmetric wave transport capabilities at finite frequencies~\cite{ma2018edge, stenull2019signatures}.
	
	To date, the experimental characterization of topological lattices has been sparse and mostly confined to specimens made of stiff materials~\cite{chen2014nonlinear, ma2018edge, paulose2015selective}. A notable exception is the recent work in~\cite{pishvar2020soft}, in which static compression and dynamic impact tests were performed on soft kagome specimens. To the authors' knowledge, no experimental studies have explored the robustness of topological properties when the lattices are mapped to curvilinear domains, resulting in a relaxation of their periodicity. 
	
	To address these gaps, this Letter investigates the mechanical response of a soft topologically polarized lattice mapped to a cylindrical domain through the engineering problem of a lattice wheel rolling on surfaces with rugged (i.e. with localized asperities) and smooth terrain profiles. We demonstrate that a topologically polarized lattice wheel has an enhanced ability to accommodate large localized deformation caused by sharp terrain profiles, thus limiting the force transfer to the axle under these types of loadings, while also maintaining satisfactory load-bearing capabilities to distributed loads. Additionally, our experiments provide further insight on the large deformation mechanical response of soft topological kagome lattices at large, enriching and complementing the results in~\cite{pishvar2020soft}.

	\section{Cylindrical Mapping of Lattice Strips}
	
\begin{figure} [!htb]
		\raggedright
		\includegraphics[scale=0.34, trim = 0cm  2cm 2cm 0cm]{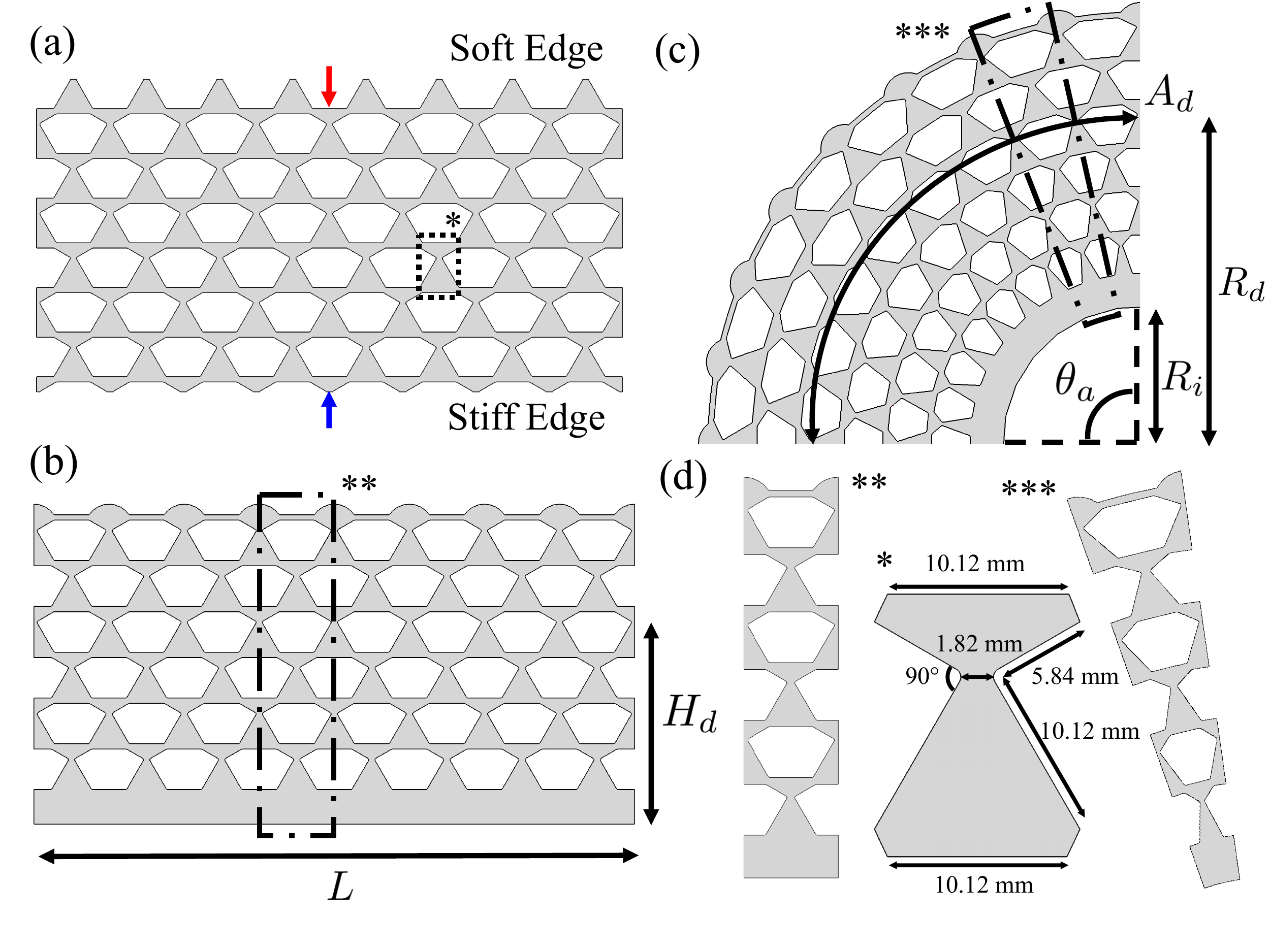}
		\caption{Topological kagome geometries before and after cylindrical mapping. (a) Primitive periodic lattice featuring polarization, whose unit cell is composed of an equilateral triangle surmounted by a non-twisted isosceles triangle. (b) Primitive periodic lattice with modified edges. (c) Sector of a lattice wheel resulting from cylindrical mapping of (b). (d) Unit cell* dimensions and supercells for (b)** and (c)***.}
		\label{CAD_Geometry}
	\end{figure}

	To realize a cylindrically mapped topological lattice, we start from the primitive kagome lattice shown in Fig.~\ref{CAD_Geometry}(a). The unit cell, as seen in Fig.~\ref{CAD_Geometry}(d), is composed of an equilateral triangle, with side lengths equal to 10.12 mm, surmounted by a non-twisted isosceles triangle, with base and side lengths equal to 10.12 mm and 5.84 mm, respectively. The triangles form an angle of  $90^o$ and are connected by a filleted ligament of width 1.82 mm at its skinniest, which functions as a non-ideal structural hinge with moderate bending stiffness. 
	
	\noindent It is appropriate to point out that the lattice formed from this unit cell tessellation features perfectly aligned bonds in the horizontal direction which, in ideal lattice conditions, corresponds to a topological phase transition~\cite{rocklin2017transformable} that lacks a formal polarization vector~\cite{kane2014topological}. Despite this, we note that the lattice still features an appreciable polarization between opposite edges~\cite{rocklin2017transformable} that can be practically exploited, and that the bond alignment relaxes upon mapping to a cylindrical domain, de facto deviating from the transitional behavior. To further motivate this geometry selection, it is also worth noting that the chosen configuration features polar symmetry upon mapping, thus yielding an inertially balanced cylindrical body.
	 
	To obtain a functional wheel design upon mapping, we slightly modify the primitive lattice design, resulting in the configuration of Fig.~\ref{CAD_Geometry}(b). On the stiff edge, we add an additional layer of triangles connected to a solid base, and on the soft edge, we add a final row of interconnected triangles with rounded caps. Topological protection guarantees that these changes will not affect the polarization of the lattice as a whole. We then subject this design to a cylindrical mapping, yielding the lattice sector with relaxed periodicity of Fig.~\ref{CAD_Geometry}(c). Key to this process is taking the length $L$ of the primitive lattice to be equal to the design arc of the cylindrical domain $A_d={\theta_a}R_d$, where ${\theta_a}$ is the selected mapping angle (e.g. $\pi/2$ for a quarter circle) and $R_d = L/{\theta_a}$ is the resulting design radius. Here, $R_d$ identifies the arc along which the dimensions of the original lattice features are preserved or, alternatively, the transition point between tangential compression ($R<R_d$) and tangential stretching ($R>R_d$), see the supplemental material for further details. Letting $H_d$ be the distance from the selected design arc features to the base of the primitive lattice, we can uniquely determine the inner radius of the circle $R_i = R_d-H_d$. With these ingredients, the mapping reads: $R = R_i + y$ and ${\theta}={\theta_a}x/L$, where $(x,y)$ are the Cartesian coordinates of the nodes defining the primitive lattice and $(R,{\theta})$ are the polar coordinates. 
	
	
	\section{Quasi-static mechanical tests}
	
\begin{figure*} [!htb]
		\raggedright
		\includegraphics[scale=0.525, trim = 0cm  3.5cm 0cm 0cm]{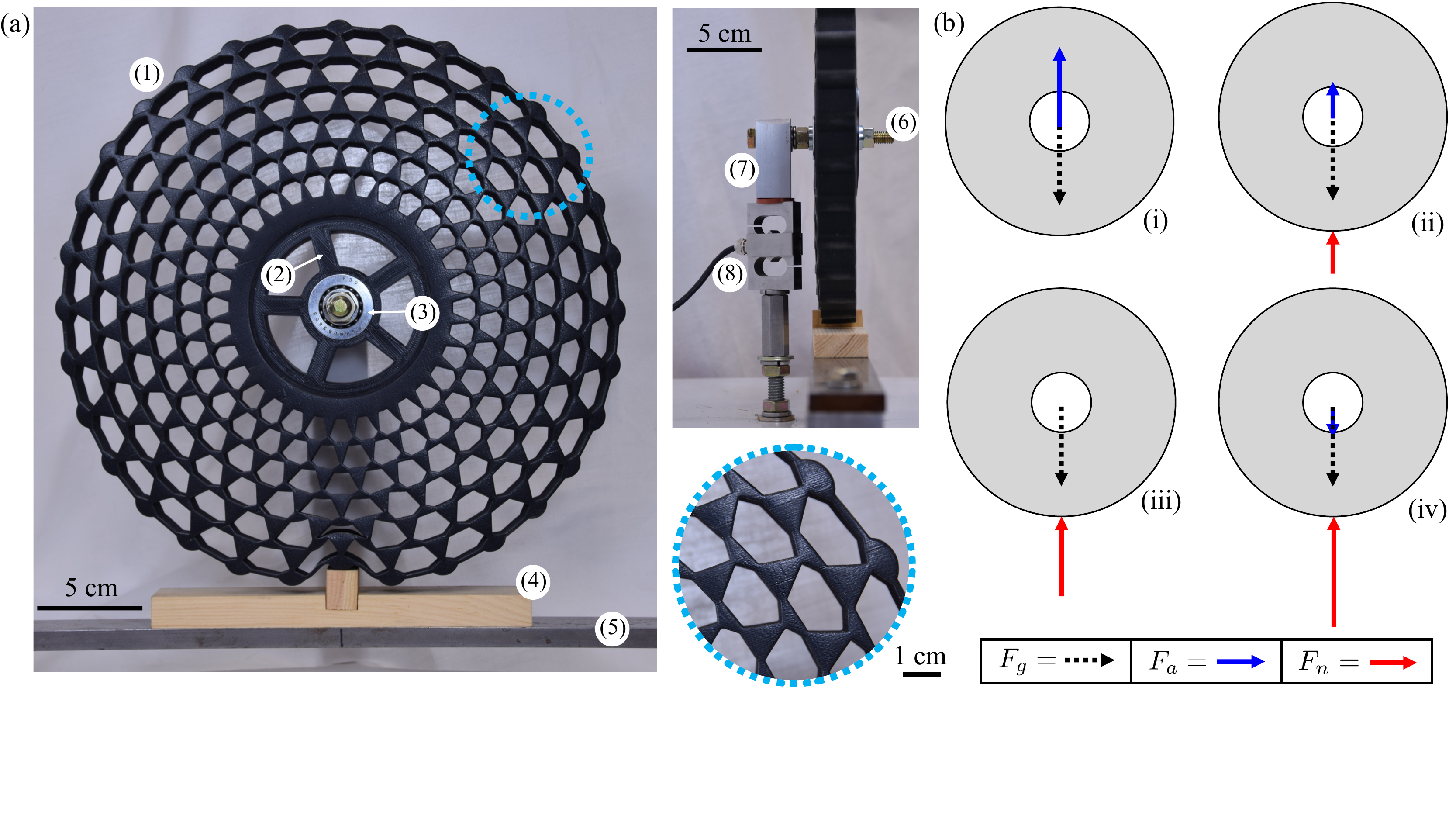}
		\caption{Experimental setup and static equilibrium considerations. (a) Front and side view of the setup. (1) 3-D printed lattice wheel, (2) rigid hub, (3) bearings, (4) terrain, (5) support track, (6) axle, (7) right angle connector, and (8) load cell. (b) Vertical force balance, where $F_g$ is the weight, $F_a$ is the force transmitted to the axle, and $F_n$ is the normal contact force at the wheel-terrain interface. Diagrams (i)-(iv) correspond to different loading stages: (i) no terrain under the wheel, (ii) small imposed displacement, (iii) displacement at which $F_a=0$, and (iv) large displacements after which the direction of $F_a$ reverses.}
		\label{Experimental_Setup}
	\end{figure*}

	We experimentally test the mechanical response of a soft topologically polarized lattice wheel obtained by mapping an extended $32{\times}6$ Fig.~\ref{CAD_Geometry}(b) lattice strip with ${\theta_a}=2{\pi}$. The lattice wheel is fabricated via PolyJet printing (Stratasys, Ltd.) from the flexible photopolymer Agilus30~\cite{stratasys2014polyjet} to an out-of-plane thickness of 25 mm. The experimental setup designed to extract the transmitted axial force due to wheel-terrain interaction is shown in Fig.~\ref{Experimental_Setup}(a). From the front we see that the  lattice wheel is mounted to a hub with bearings. Under the wheel we horizontally slide a terrain profile along a support track to cause vertical compression in the wheel. The side view shows the wheel axle supported by a load cell in charge of measuring the force on the axle. 
	
\begin{figure*} [!htb]
		\raggedright
		\includegraphics[scale=0.525, trim = 0cm  9.5cm 0cm 0cm]{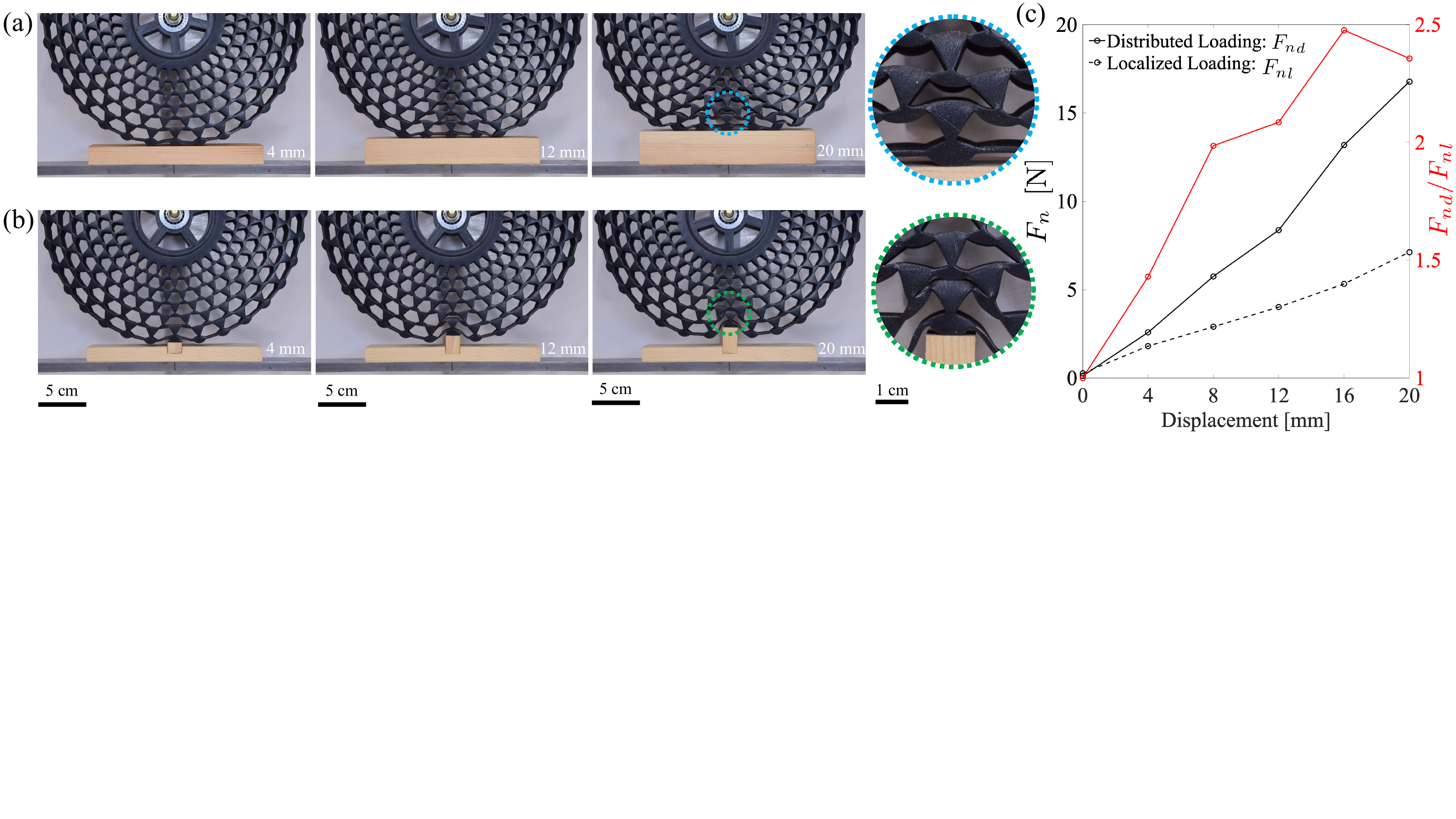}
		\caption{Experimental results. (a) Distributed loading. Shown are the 4 mm, 12 mm, and 20 mm displacements. Outlined in dashed blue is a zoomed detail of the deformed zone at the wheel-terrain interface. (b) Localized loading for the same displacements. Outlined in dashed green is a similar zoomed detail of the wheel-terrain interface. (c) Evolution of the normal force $F_n$ for the distributed and localized loadings, marked as solid and dashed lines, respectively. The red solid line corresponds to the ratio of distributed versus localized normal forces $F_{n_d}/F_{n_l}$.}
		\label{Experimental_Results}
	\end{figure*}

	We compare two types of terrain profiles: a distributed profile, shown in Fig.~\ref{Experimental_Results}(a), simulating interaction with a flat terrain that compresses the wheel evenly, and a localized profile, shown in Fig.~\ref{Experimental_Results}(b), simulating a sharp asperity in a rugged terrain. For each terrain, the imposed displacement is quasi-statically ramped from 0 to 20 mm, with the load cell measuring the axial reaction force $F_a$. While $F_a$ has direct engineering significance in monitoring the load transmitted to the axle, it is not the most convenient quantity for the sake of comparing the mechanical response to different loading scenarios. To appreciate this point, consider the force balance schematically illustrated in Fig.~\ref{Experimental_Setup}(b). Stage (i) corresponds to no terrain under the wheel (i.e. no imposed deformation). At this stage, the weight $F_g$ and the axial force $F_a$ balance each other. In stage (ii), a small terrain-imposed vertical displacement gives rise to a normal force $F_n$ at the lattice-terrain interface, thus reducing the axial reaction force needed to preserve equilibrium, such that $F_a=F_g-F_n$. Eventually, as the imposed displacement increases, stage (iii) is reached where $F_g=F_n$ and $F_a=0$. Beyond this point, in stage (iv), we observe a reversal in the direction of $F_a$. In contrast, $F_n$ increases monotonically with the loading stage, allowing for more intuitive interpretations of the measurements for comparisons. 
	
	Snapshots of successive loading stages for distributed and localized terrain are shown in Fig.~\ref{Experimental_Results}(a) and (b), respectively. The images capture the extreme deformability of the system resulting from the flexible material and from the compliance of the kagome architecture, which allows large rotations of the triangles about their connecting hinges. More importantly, for a highly localized loading, as in Fig.~\ref{Experimental_Results}(b), the response features large deformation concentrated in the neighborhood of the contact point, which sharply decays when moving into the lattice bulk. This behavior is consistent with the expectations for a polarized lattice loaded at the floppy edge, demonstrating that the polarization attributes are robust against cylindrical mapping endowing the wheel configuration with a mechanism to absorb sharp asperities in the terrain profile.
	
	To quantify this effect, we consider six loading stages between 0 and 20 mm of imposed displacement $d_i$. At each stage, we infer the normal force, labeled $F_{n_d}$ and $F_{n_l}$ for the distributed and localized loading, respectively. These quantities are plotted in Fig.~\ref{Experimental_Results}(c) as solid and dashed lines, respectively. We observe a distinct trend of divergence between the loading scenarios, well captured by the ratio $F_{n_d}/F_{n_l}$, plotted in Fig.~\ref{Experimental_Results}(c), that grows with increased $d_i$. This trend in $F_{n_d}/F_{n_l}$ indicates the coexistence of two dichotomic properties in the response: a local softness against localized loads and a global stiffness against distributed ones. This property, endowed to the design solely through its geometric construction, gives the lattice wheel the ability to absorb sharp puncturing loadings without resorting to extreme material softness, while simultaneously preserving an overall load-bearing capability. To assess the significance of this result, it is necessary to determine to what extent these properties can be attributed to the specific deformation characteristics of the lattice architecture, as opposed to being a generic feature of any circular body subjected to imposed displacements at their boundaries. To this end, we resort to finite element simulations to compare the response of the lattice wheel against that of a solid wheel made of the same material.
	
    \section{Simulation-Enabled Design Comparisons}
    
	We simulate the geometrically nonlinear static response of our experimental lattice wheel and compare it to that of a solid wheel with identical inner and outer radii. The nonlinear finite element simulations are performed in Abaqus via the Abaqus/Standard solver. For both configurations, the material properties are selected to mimic the behavior of Agilus30, used in our experiments, which is modeled as linear elastic~\cite{abayazid2020material} with Young's modulus $E=0.65$ MPa~\cite{dykstra2019viscoelastic,mirzaali2020multi}, Poisson's ratio ${\nu}=0.34$~\cite{han2018structural}, and density ${\rho}=1.15$ $\mathrm{g/cm^3}$~\cite{stratasys2014polyjet}. Further discussion regarding the material property selection and the adoption of a linear elastic constitutive model is given in the supplemental material. The solid is discretized with plane-strain, hybrid, quadratic, six-node, triangular elements (Abaqus element type CPE6H), with density determined through mesh refinement studies. To mimic closely the experimental conditions, a two stage loading process is simulated. The first stage applies a gravitational load and the second imposes a vertical displacement at the base of the wheel. The distributed displacement is simulated by pushing a flat rigid body against the base of the wheel with hard contact enforced at the interface. The localized displacement is simulated by imposing a displacement to a portion of the wheel edge equal in width to that of the localized terrain profile (i.e. 15 mm). To validate the deformation fields and the normal contact forces obtained via numerical simulations, we compare them against the displacement fields reconstructed from experimental data acquired using digital image correlation (DIC)~\cite{jones2018good, turner2015digital} and the normal forces found through the load cell data, respectively. Details of the DIC tests and of the resulting comparisons are provided in the supplemental material.
	
	\begin{figure*} [!htb]
		\raggedright
		\includegraphics[scale=0.525, trim = 0cm  3cm 0cm 0cm]{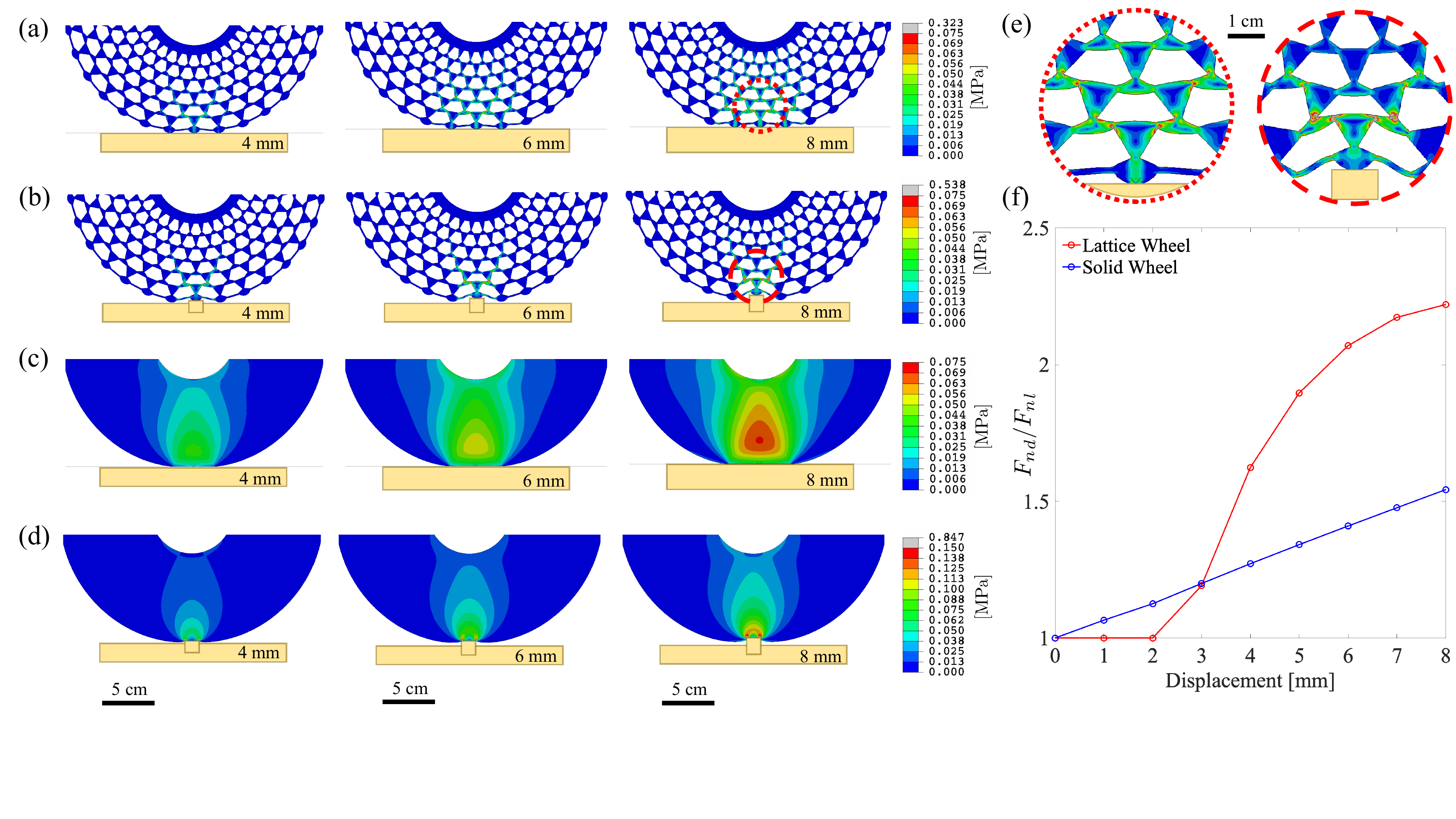}
		\caption{Numerical simulation results. For (a)-(d) the displacements shown are 4 mm, 6 mm, and 8 mm. The color maps correspond to von Mises stress in MPa. (a) Distributed loading applied to the lattice wheel. (b) Localized loading applied to the lattice wheel. (c) Distributed loading applied to the solid wheel. (d) Localized loading applied to the solid wheel. (e) Dotted red line and dashed red lines show zoomed details of distributed and localized loading at 8 mm at the wheel-lattice interface, respectively. (f) Evolution of the normal force ratios $F_{n_d}/F_{n_l}$ for the lattice (red) and solid (blue) wheels.}
		\label{FEM_Results}
	\end{figure*}
	
	We peform four simulations, encompassing the two loading scenarios for each wheel configuration. For each case, the imposed vertical displacement $d_i$ is ramped from 0 to 8 mm. Three successive snapshots of the deformation process for each case are shown in Fig.~\ref{FEM_Results}(a)-(d), where the color map is proportional to the von Mises stress. We also compute the total vertical reaction force at the wheel axle $F_a$ from the resultant of the stresses at the inner circumferential edge. To match the experiments, we work with the normal force $F_n = F_g - F_a$, where $F_g$ is computed after the gravitational load is applied. By monitoring the value of $F_n$ throughout the loading process for each of the four simulations, we compute $F_{n_d}/F_{n_l}$ for both wheels. This quantity is plotted for the lattice and solid wheels in Fig.~\ref{FEM_Results}(f) as red and blue curves, respectively. For the solid wheel, $F_{n_d}/F_{n_l}$ grows linearly with $d_i$, but remains below 1.5, suggesting a minor difference in transmitted force between the two loading scenarios. This is further substantiated by the von Mises color maps in Fig.~\ref{FEM_Results}(c) and (d), which show large stress penetration into the bulk in both cases, albeit slightly more pronounced for the distributed loading--an expected feature given that a flat object engages more points on the boundary compared to a sharp indenter. In light of this, the lattice wheel exhibits a noticeably more dramatic evolution of $F_{n_d}/F_{n_l}$ compared to the solid wheel. Initially, for the lattice wheel, $F_{n_d}/F_{n_l}$ remains flat around 1, suggesting that, for small $d_i$, the localized and distributed loads result in stress fields with similar penetration depths. This behavior is caused by the distributed and localized loads engaging the boundary in identical ways at small $d_i$ (i.e. both types of loads engage only the bottommost unit cell directly below the axle). However, for sufficiently large $d_i$, the difference in the stress fields, and ultimately the transmitted axial force, induced by the two loading scenarios become progressively more accentuated, as observable by comparing Fig.~\ref{FEM_Results}(a) and (b). Additional mechanistic rationale for this behavior, which is intrinsically rooted in the ability of floppy edges to host localized soft modes, can be found by considering the different deformation fields activated by the two loading scenarios. The distributed loading promotes a horizontal alignment of the bonds, triggering stiff local structural features that accumulate stresses and strains~\cite{kane2014topological, sun2012surface, lubensky2015phonons}. In contrast, the localized loading activates large bending of the hinges and relative rotations of the triangles, thus relaxing the bond alignment landscape and triggering softer deformation mechanisms. Nevertheless, we also note that, at larger $d_i$, $F_{n_d}/F_{n_l}$ plateaus, suggesting that, at these advanced stages, additional deformation mechanisms, including densification and friction of the solid triangles, begin to accompany or dominate the bending of the hinges. To summarize these observations, the lattice wheel possesses an enhanced dichotomous character, whereby it responds softly to sharp penetrating objects while exhibiting a significantly stiffer response against flat ones. Looking at this property through the engineering prism of wheel-axle interaction, the coexistence of soft and stiff mechanisms enables absorption of terrain asperities minimizing spikes in stress transfer to the axle, while maintaining satisfactory global stiffness to bear payload.

	 \section{Concluding Remarks}
	 
	 In conclusion, this work has provided a first look into the mechanics of soft topologically polarized kagome lattices subjected to cylindrical mapping through the engineering problem of a lattice wheel rolling on an irregular surface. We have revealed dichotomic properties in the form of local edge softness against localized indenting objects, accompanied by satisfactory stiffness against flat surfaces. Importantly, this behavior is robust against perturbations in the lattice geometry (e.g. a layer of outer cells being peeled away), endowing the wheel with robustness against damage and material degradation. These features make the proposed design ideal for applications where locomotion on remote rugged terrains is essential, including the design of military vehicles and rovers for space exploration, scientific sampling, and disaster relief. These opportunities offer a tangible bridge between the concepts of topological mechanics and technological applications~\cite{gonella2020non}, thus contributing to a broader adoption of engineering solutions enabled by mechanical metamaterials at large.
	 
	  \section{Acknowledgments}
	  The authors acknowledge the support of the National Science Foundation (NSF Grant No. EFRI-1741618). The authors would like to thank Lauren Linderman for graciously allowing extended used of experimental equipment, Xiaoming Mao for useful discussions regarding topological mechanics, and David Zunker for his help on the experimental work.

	\bibliography{bib}
		
	\newpage

	\section{Supplemental Material}

	\subsection{Distortion Due to Cylindrical Mapping}
	
	Comparison of the cylindrically mapped supercell in Fig.~\ref{CAD_Geometry}(d) to that of the primitive supercell reveals the tangential distortion induced by cylindrical mapping in the form of tangential stretching at $R>R_d$ and tangential compression at $R<R_d$. To quantify the rate of tangential distortion, we consider the chord length $c$ as a function of radius along a sector spanning  a given angle $\theta$. Given that $c=2R\sin({\theta}/2)$ we see that two chords at an inner radius $R_1$ and outer radius $R_2$ ($R_1<R_2$) are related as $c_2/c_1 = R_2/R_1$. Using this relation the tangential stretching or compression of features (e.g. the width of a hinge) away from $R_d$ can be approximately calculated. 
	
	\subsection{Agilus30: Constitutive Model and Material Properties}
	
\begin{figure} [!htb]
		\raggedright
		\includegraphics[scale=0.725, trim = 0cm  4cm 10cm 0cm]{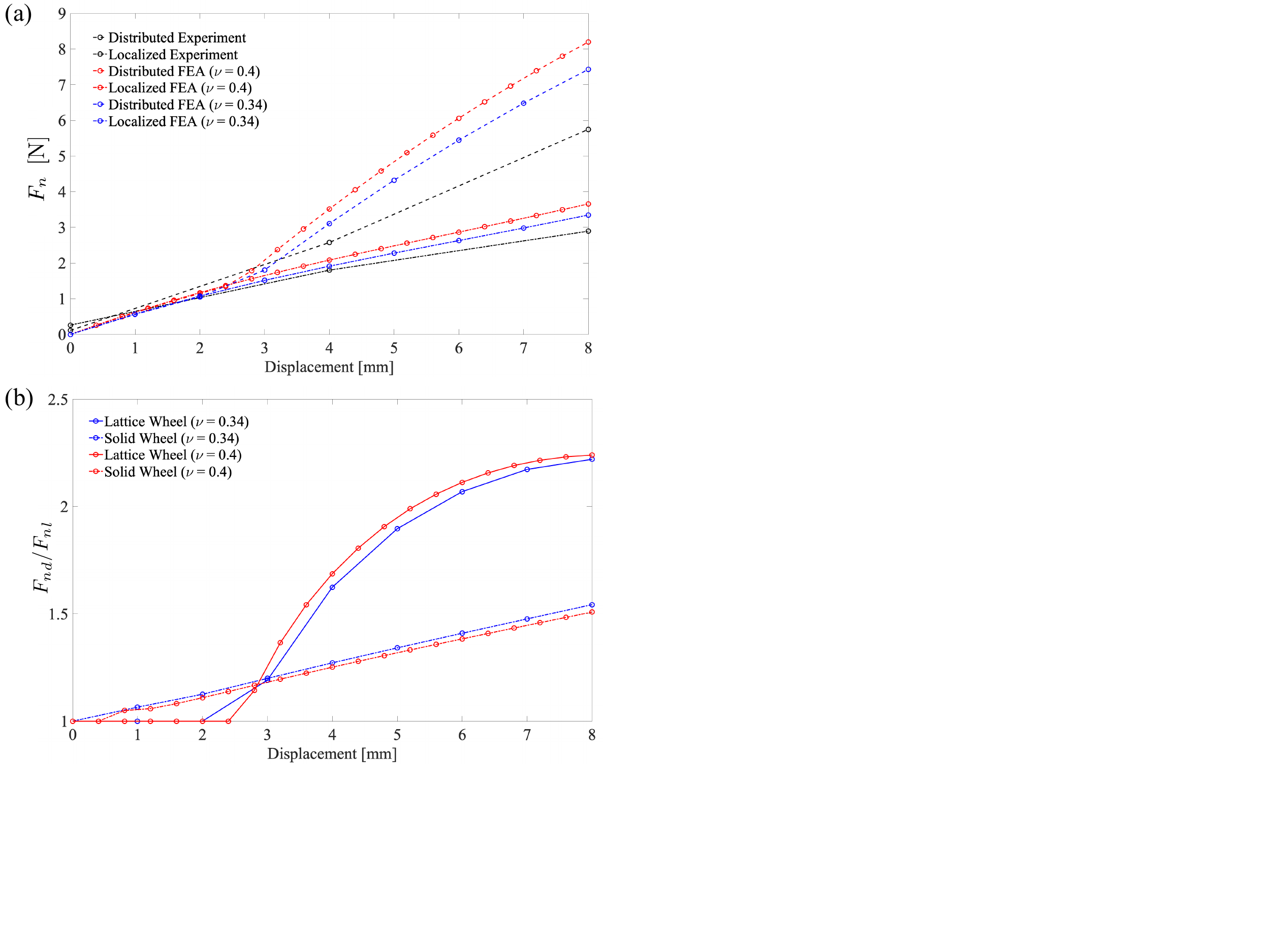}
		\caption{Investigation of the effect of changing the Poisson's ratio. (a) Comparison of normal force values obtained for the lattice wheel including: experimental data, finite element simulation with ${\nu}=0.4$, and finite element simulation with ${\nu}=0.34$. (b) Evolution of the normal force ratios $F_{n_d}/F_{n_l}$ for the lattice and solid wheels with different Poisson's ratios.}
		\label{Poisson_Ratio_Investigation}
	\end{figure}

	In the finite element simulations the material of both the lattice wheel and the solid wheel is modeled as Agilus30 (the experimental printing material), using a linear elastic constitutive model with Young's modulus $E=0.65$ MPa, Poisson's ratio ${\nu}=0.34$, and density ${\rho}=1.15$ $\mathrm{g/cm^3}$. The selection of a linear elastic constitutive model is motivated by a recent comprehensive study~\cite{abayazid2020material} focusing on accurately modeling PolyJet elastomers, including Agilus30, in the attempt to address numerous discrepancies between reported material models in previous literature. The work reveals that the mechanical response of Agilus30 has a strong dependency on strain rate and can also depend on build orientation. Importantly, their material testing data shows that, as the loading rate tends to the quasi-static limit, the response, especially for low strains, can be approximated as linear elastic. Given that our finite element analysis does not exceed strains of 10\% and that it is intended to capture the experimental conditions in which loading was applied quasi-statically, we adopt a linear elastic constitutive model. To select the material properties the elastic modulus inferred from the data in~\cite{abayazid2020material} is compared to two other literature values~\cite{dykstra2019viscoelastic,mirzaali2020multi} which all show good agreement at a value of approximately $E=0.65$ MPa. The Poisson's ratio poses as a more troubling value to verify because, to the authors' knowledge, it is only reported in~\cite{han2018structural} as ${\nu}=0.34$, a value that is below the Poisson's ratio classically expected for rubber-like materials. To address this concern, we test the effects of assuming a higher Poisson's ratio of ${\nu}=0.4$ in the numerical simulations and compare it to the results found with ${\nu}=0.34$ as well as to the experimental results for the lattice wheel, as shown in Fig.~\ref{Poisson_Ratio_Investigation}(a). These investigations reveal that an increase in ${\nu}$ actually decreases the agreement with the experimental observations, furthermore, it produces a minimal difference in the evolution of $F_{n_d}/F_{n_l}$ for both the lattice and solid wheel, ultimately not altering the relative comparison between the two types of terrains, as shown in Fig.~\ref{Poisson_Ratio_Investigation}(b). For these reasons the reported literature value of ${\nu}=0.34$ is accepted. The density of Agilus30 is taken to be that reported by the manufacturer~\cite{stratasys2014polyjet}.
	
	\subsection{Digital Image Correlation}
	In order to verify the numerical simulations digital image correlation (DIC) is performed to infer displacements from experiments. The following paragraphs report all the hardware used and the analysis techniques followed as recommended by the International DIC Society (iDICs) ~\cite{jones2018good}.
	
\begin{figure} [!htb]
		\raggedright
		\includegraphics[scale=0.55, trim = 0cm  9cm 10cm 0cm]{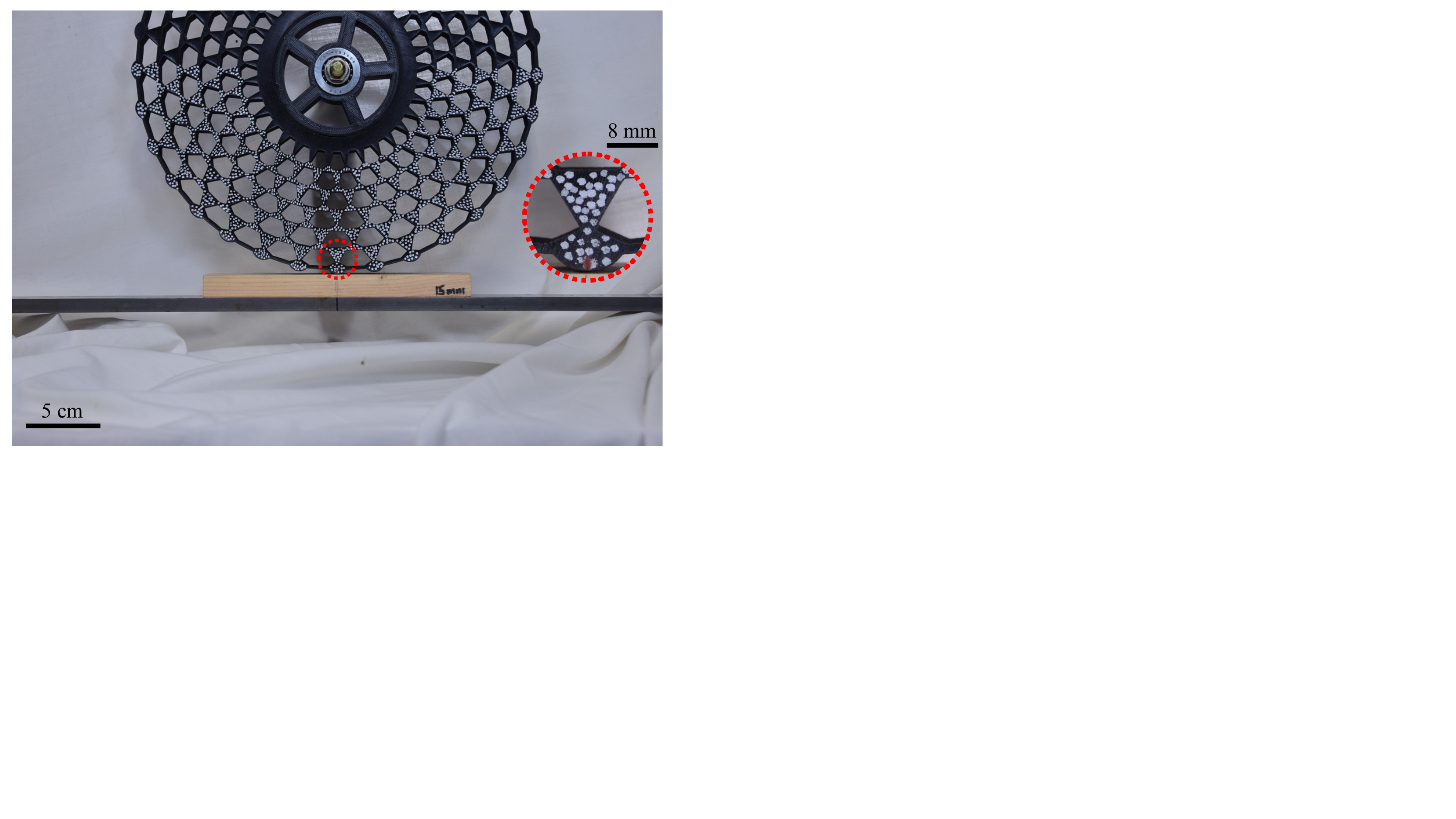}
		\caption{Field of view for digital image correlation. The dotted red line shows a zoomed detail of the speckle pattern created with a silver, oil-based, medium point Sharpie marker.}
		\label{DIC_FOV}
	\end{figure}

	Physical Experimentation: The camera is a Nikon D5300 DSLR with an image resolution set to 2992x2000 px. The lens is a Nikon AF-S DX Nikkor 18-200mm f/3.5-5.6G ED VR II standard zoom lens. During data collection, the zoom is set to 50 mm and the focus to approximately 1 m which is equivalent to the stand-off distance. The shutter speed of the camera is set to 1/25 (sec), the aperture to f/8, and the ISO to 100. The field of view is shown in figure Fig.~\ref{DIC_FOV}. No image scaling is performed. The image acquisition rate is not applicable because images of the static configurations are taken after the terrain has been placed. The speckle pattern is applied with a silver, oil-based, medium point Sharpie marker, which results in speckles with 1 mm diameters. The speckle pattern is shown in Fig.~\ref{DIC_FOV} in the image bordered by a red dashed line (the red dot was used for alignment). 
	
	Analysis: The DIC software used for analysis is DICe~\cite{turner2015digital} created by Sandia National Laboratories. During analysis Gauss image filtering is applied. The subset size is set to 29 px (or 4.2 mm)  and the step size to 18 px (or 2.6 mm). The shape functions for translation and rotation are activated. The noise floor is calculated at {$\pm$} 0.146 mm and the out-of-plane error is estimated to be at most 0.5 mm.
	
	\subsection{Numerical Verification}
\begin{figure*} [!htb]
		\raggedright
		\includegraphics[scale=0.7, trim = 0cm  2.8cm 2cm 0cm]{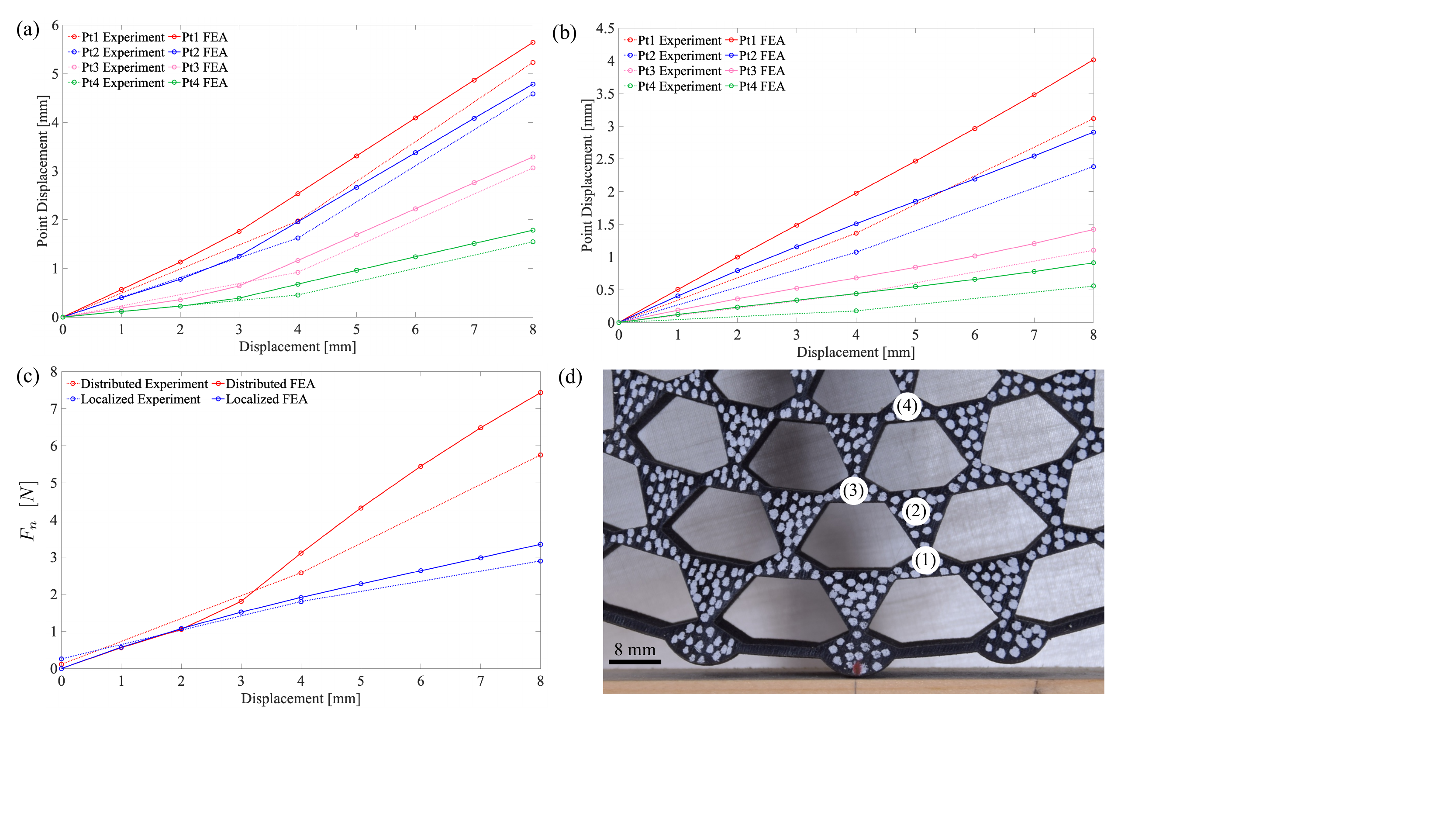}
		\caption{Validation of numerical simulations by comparison against experimental data. (a) Comparison of displacements of selected tracking points for the distributed loading. (b) Comparison of displacements of selected tracking points for the localized loading. (c) Comparison of normal force inference between the distributed and localized loadings. (d) Enhanced view of the lattice wheel to show points 1-4 that were tracked during digital image correlation.}
		\label{Experimental_Verification}
	\end{figure*}

	The accuracy of the numerical simulations is validated by comparison of its results to the following two experimental quantities: the normal force $F_n$ found from the load cell data and four selected nodal displacements measured through digital image correlation (DIC) that are shown in Fig.~\ref{Experimental_Verification}(d). Fig.~\ref{Experimental_Verification}(c) shows the comparison of $F_n$ and Fig.~\ref{Experimental_Verification}(a) and (b) show the comparison of the displacement field for the distributed and localized loading cases, respectively. Inspection of Fig.~\ref{Experimental_Verification}(a) and (b) reveals that the displacement fields between the experiment and numerical simulations shows good agreement, with deviations between the two never exceeding 1 mm. Inspection of Fig.~\ref{Experimental_Verification}(c) shows satisfactory qualitative agreement between normal forces for the numerical simulations and experiment.
	
	\newpage
	
\end{document}